\documentclass[prl,preprint,showpacs,amsmath,amsfonts,endfloats]{revtex4}

\usepackage{bm}
\usepackage{graphicx}

\begin{document}

\title{Fractal templates in the escape dynamics of trapped ultracold atoms}

\author{Kevin A.~Mitchell}

\affiliation{School of Natural Sciences, University of California,
Merced, California, 95344}

\author{Daniel A.~Steck}

\affiliation{Oregon Center for Optics and Department of Physics, 1274 University of Oregon, Eugene, Oregon 97403-1274}

\date{\today}

\begin{abstract}
We consider the dynamic escape of a small packet of ultracold atoms
launched from within an optical dipole trap.  Based on a theoretical
analysis of the underlying nonlinear dynamics, we predict that fractal
behavior can be seen in the escape data.  This data would be collected
by measuring the time-dependent escape rate for packets launched over
a range of angles.  This fractal pattern is particularly well resolved
below the Bose-Einstein transition temperature---a direct result of
the extreme phase space localization of the condensate.  We predict
that several self-similar layers of this novel fractal should be
measurable and we explain how this fractal pattern can be predicted
and analyzed with recently developed techniques in symbolic dynamics.
\end{abstract} 

\maketitle

\begin{figure}
\includegraphics[width = 3.5in]{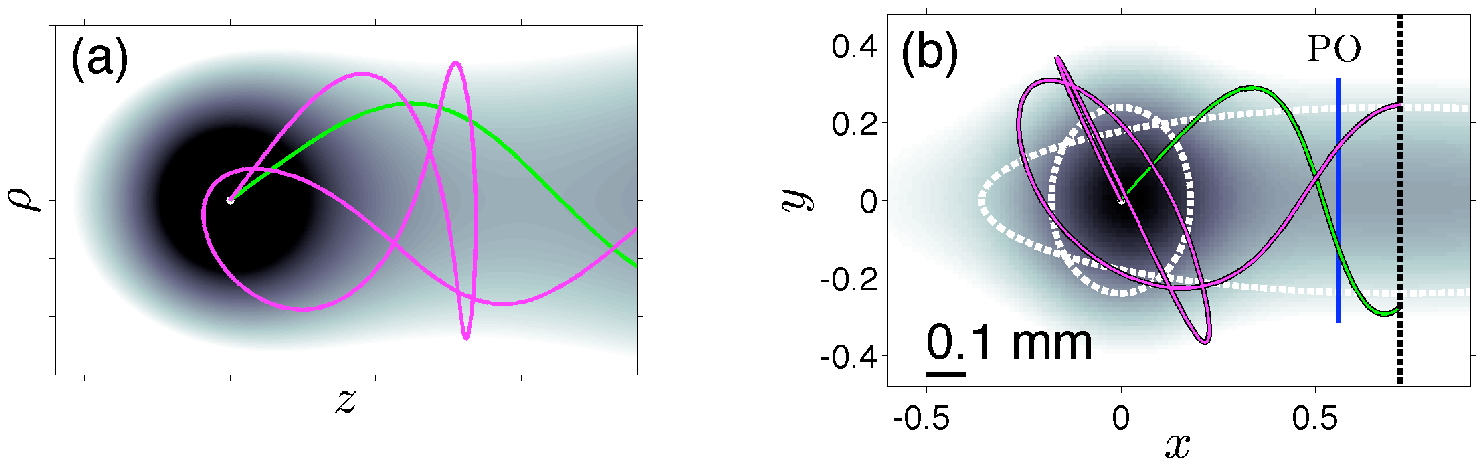}

\caption{\label{fpotentials} a) The potential energy for a hydrogenic
electron in applied parallel electric and magnetic fields. Two
ionizing trajectories are shown.  b) The double-Gaussian potential.
Gaussian widths are indicated by the dashed ellipses.}
\end{figure}

Chaotic escape is a widespread transport process underlying such
diverse phenomena as conductance through ballistic
microstructures~\cite{Taylor03}, emission from deformed micro-disk
semiconductor lasers~\cite{Gmachl98}, molecular scattering and
dissociation~\cite{Molecules}, celestial transport~\cite{Celestial},
and atomic ionization~\cite{ChaoticIonization,Mitchell04}.  We have
been particularly motivated by the chaotic ionization of hydrogen in
applied parallel electric and magnetic fields, for which a recent
theoretical analysis
\cite{Mitchell04} predicts that the time-spectrum for ionization
will display a \emph{chaos-induced train} of electron pulses.  This
prediction is based on classical ionizing trajectories
(Fig.~\ref{fpotentials}a), which propagate from the nucleus into the
ionization channel, via the Stark saddle.  These trajectories exhibit
fractal self-similarity, which is reflected in the pulse train.  This
prediction has been recently confirmed by full quantum
computations~\cite{Turker}.  However, the experimental observation of
these chaos-induced pulse trains remains unrealized.

In this Letter, we propose an alternate physical system---the escape
of ultracold atoms in specially tailored optical dipole traps---that
exhibits similar escape dynamics.  However, we show that the
flexibility and control afforded by cold atoms, especially in
engineering the initial state, should permit the {\em direct} imaging
of fractals in the escape dynamics, including novel self-similar
features.  Furthermore, we show how this self-similarity can be
analyzed using a recently developed symbolic formalism.
Fundamentally, these fractals result from homoclinic tangles
\cite{Wiggins92}---a general mechanism for phase space transport.
Hence this Letter suggests that cold atoms could serve as a unique
high-precision experimental probe of this mechanism.  Finally, the
cold atom experiments discussed here are readily feasible with
present-day experimental configurations and should prove easier to
realize than the previously mentioned ionization experiments.

Recent experiments from the Raizen~\cite{Raizen01} and
Davidson~\cite{Davidson01} groups have made first steps along these
lines.  They independently measured the {\em long-time} survival
probability for ultracold atoms escaping through a hole in an optical
billiard, demonstrating the distinction between regular and chaotic
escape dynamics.  In contrast, our Letter focuses on the short- to
intermediate-time dynamics, where fundamentally distinct phenomena,
such as fractal self-similarity, are predicted to appear.

{\em The double-Gaussian trap:} We consider a dipole potential
consisting of two overlapping Gaussian wells
\begin{align}
V(x,y) & = - V_1 \exp( - [(x/\sigma_{1x})^2 + (y/\sigma_{1y})^2]/2)
\nonumber \\
& - V_2 \exp( - [((x-x_2)/\sigma_{2x})^2 + (y/\sigma_{2y})^2]/2),
\end{align}
as shown in Fig.~\ref{fpotentials}b.  This potential can be created by
two red-detuned, far-off-resonant Gaussian beams; atomic motion can be
further restricted transverse to the $xy$-plane by a uniform laser
sheet.  Here, we take $\sigma_{1x}=0.18$, $\sigma_{1y}=0.24$,
$\sigma_{2x}= 1.08$, $\sigma_{2y}=0.24$, $x_2 = 0.72$ (measured in
millimeters), and $V_1 = V_2 = 35.5$ (measured in recoil energies
$E_\mathrm{r} = \hbar^2 k_\mathrm{\scriptscriptstyle
L}^2/2m_\mathrm{\scriptscriptstyle Rb} = h \cdot 3.77 \;
\mathrm{kHz}$, where $\lambda = 2 \pi / k_\mathrm{\scriptscriptstyle
L} = 780.2 \;
\mathrm{nm}$ for the D$_2$ transition of $^{87}$Rb.)
The double-Gaussian potential shares several features in common with
the hydrogen potential in Fig.~\ref{fpotentials}a.  The ``primary''
Gaussian centered at the origin is analogous to the Coulomb well; the
elongated ``secondary'' Gaussian on the right is analogous to the
ionization channel; and the saddle connecting the two Gaussian wells
is analogous to the Stark saddle.  We are interested in the transport
of atoms from the primary well into the secondary well.
Fig.~\ref{fpotentials}b shows two representative trajectories that
move away from the origin with initial speed 4.12 cm/s, pass over an
unstable periodic orbit (PO) near the saddle, and then strike a
resonant laser sheet (the vertical dashed line.)  This sheet forms a
{\em detection line} that serves both to image the escaping atoms and
to scatter them out of the trap, preventing their return into the
primary well.

\begin{figure}
\includegraphics[width = 3.5in]
{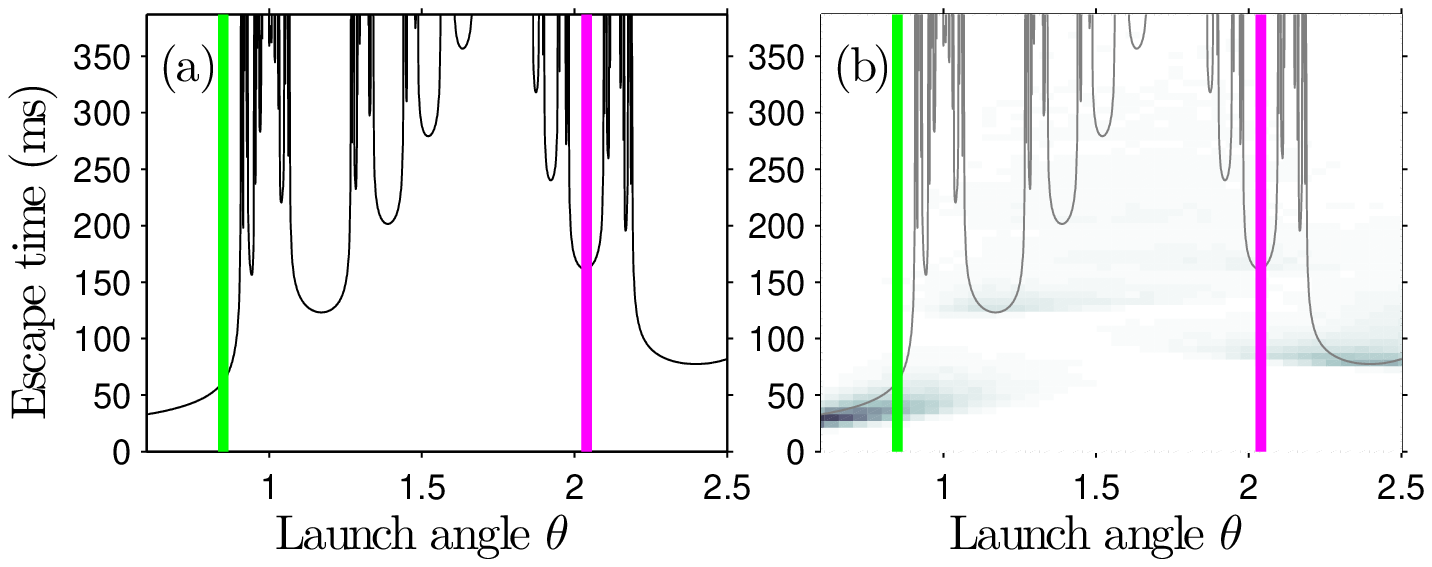}
\caption{\label{fdg_etplot} a) The escape-time plot.   Vertical lines denote $\theta$ for the trajectories in Fig.~\ref{fpotentials}b.
b) The shading shows the escaping flux from an initial packet of size
500 times Planck's constant.}
\end{figure}

Figure~\ref{fdg_etplot}a plots the time for a trajectory with initial
speed 4.12 cm/s (energy -14.9 D$_2$ recoils) to move from the origin
to the detection line, as a function of launch angle $\theta$,
measured relative to the positive $x$ axis.  The resulting {\em
escape-time plot} is highly singular, with numerous icicle-shaped
regions whose edges tend toward infinity.  These ``icicles'' exhibit a
self-similar fractal pattern.  Such patterns also occur in the chaotic
ionization of hydrogen and are characteristic of chaotic escape and
scattering.

\begin{figure}
\includegraphics[width = 3.25in]{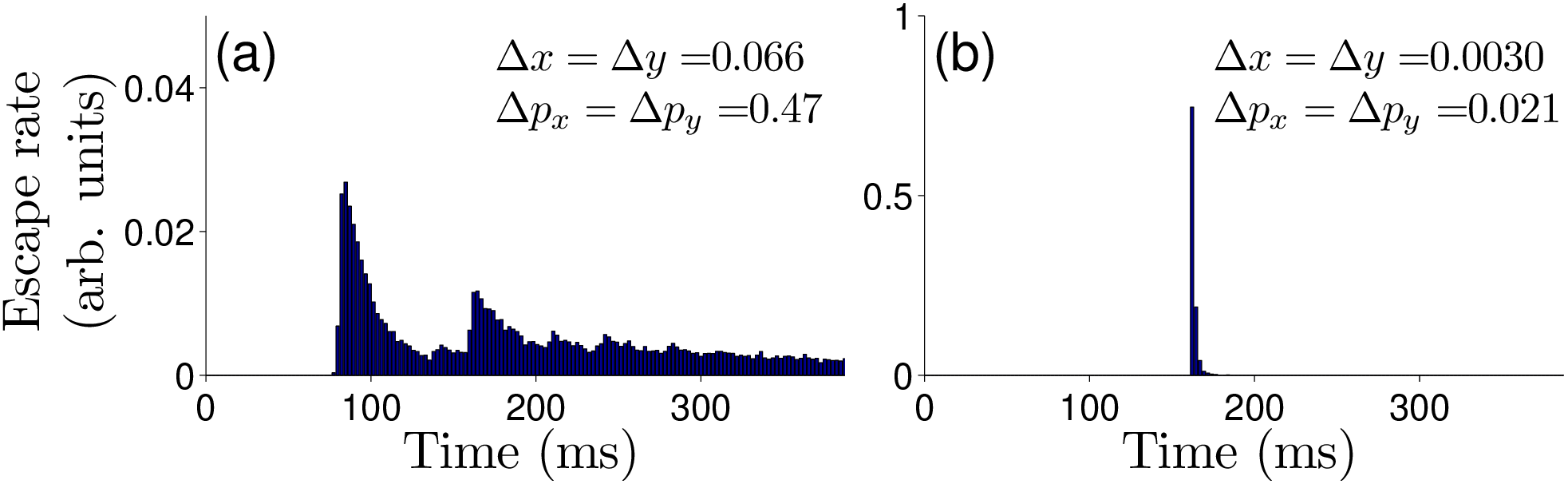} 
\caption{\label{fsingle_pulsetrain} The flux of atoms reaching the
detection line for two different Gaussian ensembles launched at
$\theta = 2.04$.  The position and momentum widths of the initial
ensembles are given in units of mm and recoil momenta ($p_\mathrm{r} =
\hbar k_\mathrm{\scriptscriptstyle L} = m_\mathrm{\scriptscriptstyle Rb}~\cdot~5.88 \;
\mathrm{mm}/\mathrm{s}$ for the $^{87}$Rb D$_2$ transition.)}
\end{figure}

{\em A proposed experiment to measure self-similar patterns in the
escape-time plot:} We consider a small Gaussian packet of ultracold
atoms launched from the origin with speed 4.12 cm/s and $\theta =
2.04$ (the right line in Fig.~\ref{fdg_etplot}a).  The subsequent flux
of atomic trajectories at the detection line is then computed as a
function of time.  Fig.~\ref{fsingle_pulsetrain}a shows this flux for
an initial thermal packet that occupies a phase space area 500 times
Planck's constant in both the $x$ and $y$ degrees of freedom.
Fig.~\ref{fsingle_pulsetrain}b, on the other hand, uses a packet that
occupies a single Planck cell, appropriate for a pure dilute
Bose-Einstein condensate (BEC) in the regime of negligible
interactions.  The condensate packet closely follows the trajectory in
Fig.~\ref{fpotentials}b, exiting as a sharp pulse at 160 ms, near the
bottom of the $\theta = 2.04$ icicle in Fig.~\ref{fdg_etplot}a.  The
thermal packet also produces a pulse at $160$ ms, but its larger phase
space extent populates neighboring icicles, thereby producing the
additional pulses in Fig.~\ref{fsingle_pulsetrain}a.

By repeating the preceding computation for different launch angles of
the thermal packet, we obtain the aggregate data in
Fig.~\ref{fdg_etplot}b, where the shading records the atomic flux as a
function of arrival time and the packet's launch angle.
Fig.~\ref{fsingle_pulsetrain}a then corresponds to the vertical slice
through Fig.~\ref{fdg_etplot}b at $\theta = 2.04$.  The thermal data
appear as a blurred version of the sharp escape-time plot.  For
example, the left and right icicles are associated with prominent dark
patches, and in between a few wispy patches can be associated with the
bottoms of other icicles.  Overall, however, the intricate icicle
structure is poorly resolved by the thermal data.

\begin{figure}
\includegraphics[width = 3.6in]{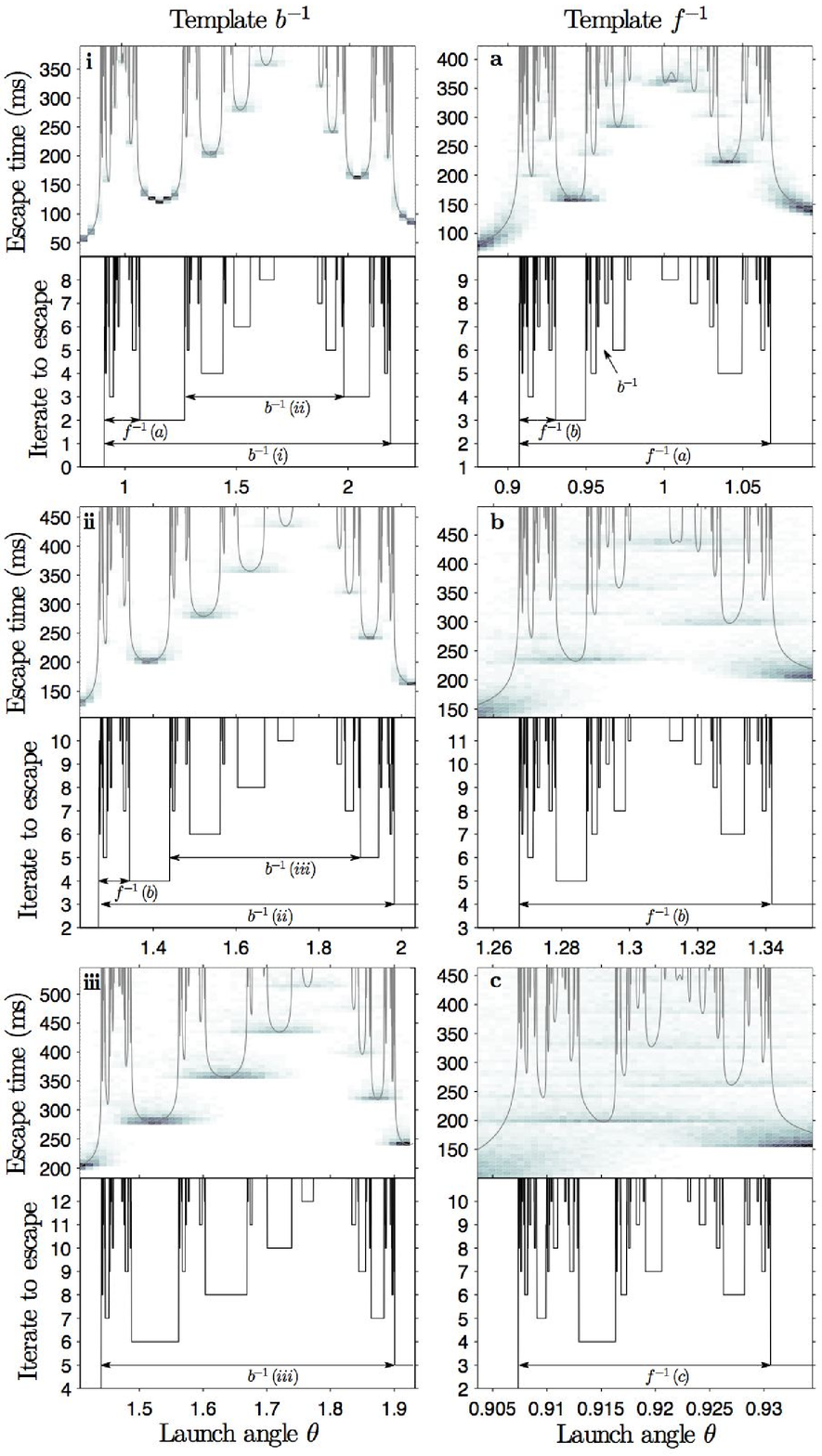}
\caption{\label{ftemplates} Escape time  over six different
angular intervals.  Each escape-time plot is matched below by its
corresponding discrete-escape-time plot.}
\end{figure}

A remarkable increase in resolution occurs below the BEC transition,
shown in Fig.~\ref{ftemplates}i.  Many icicles are now clearly
resolved, a direct consequence of the high phase-space localization
afforded by the condensate.  This increase in resolution prompts us to
look deeper into the fractal, with the expectation that we can
directly measure its fine-scale structure.

{\em Self-similarity of the escape-time data:} The first column of
Fig.~\ref{ftemplates} shows data plotted for three distinct intervals
of the launch angle.  (We concentrate at present on the upper plot in
each pair.)  The three plots look remarkably similar, and icicles in
one plot can be identified with icicles in the other two.  In fact,
this pattern of icicles occurs throughout the escape-time plot and on
all scales.  (See below.)  One of the principal observations of this
Letter is that these structures are resolved by the overlaid BEC
data.  The icicles look progressively more blurred as we move down the
column because the interval width is decreasing and the escape time
is increasing.

The pattern in the first column of Fig.~\ref{ftemplates} is not the
only repeated pattern.  The second column of Fig.~\ref{ftemplates}
shows another pattern that also occurs on all scales throughout the
{\em same} escape-time plot.  As we will see, many such repeated
patterns, or {\em templates}, exist in the escape-time plot.  Within a
given template, all other templates can be found on smaller scales.
That is, each template occurs as a subtemplate of every other template
in an infinitely recursive nesting.  Our computations predict that
several nested layers will be experimentally visible.

Experimentally, the observation of these phenomena will not be easy
but certainly feasible.  With a 1.06~$\mu$m laser, the above trap
geometry can be realized with about 80~W of power.  With this
detuning, a $^{87}$Rb atom has at most a 4\% probability of
spontaneous scattering over a half second.  Acceleration of the atoms
to the initial velocity of 4.12~$\mathrm{cm}/\mathrm{s}$ is easily
accomplished; for example, a chirped, one-dimensional optical lattice
of 785~nm light with 50~mW of single-beam power and beam radius  
$w_0=100\;\mu$m can accelerate the
atoms in about 1.6~ms with negligible ($<1\%$) probability of
spontaneous scattering and heating due to energy-band transitions.
The primary difficulty is the subrecoil initial conditions required,
even in the thermal case.  However, a standard expanded BEC should
suffice.

\begin{figure}
\includegraphics[width = 3.25in]{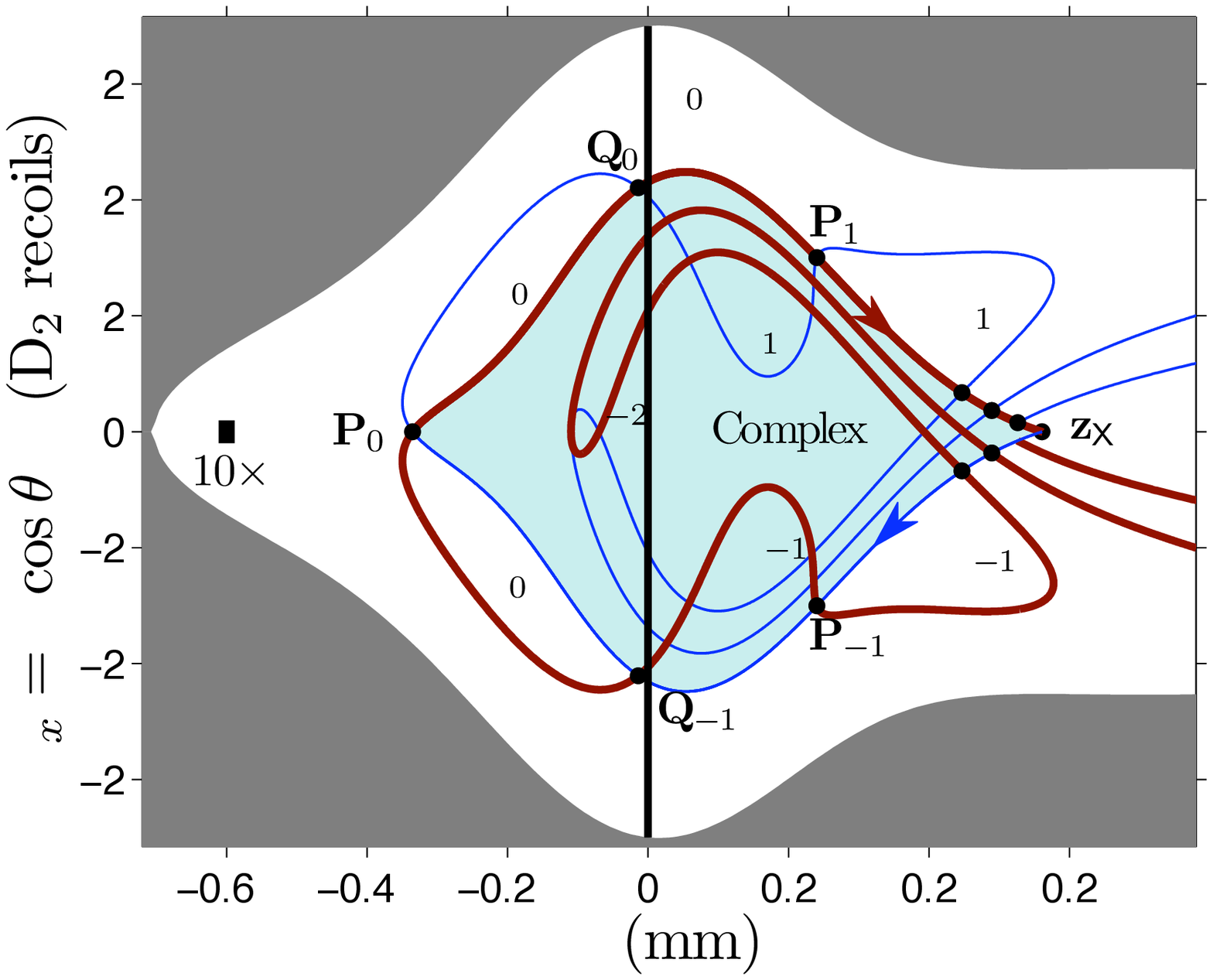}
\caption{\label{fsos} The surface of section, showing the unstable fixed point $\mathbf{z}_\mathsf{X}$ and its associated homoclinic tangle.  The left rectangle has  area ten times Planck's constant.}
\end{figure}

\textit{Theoretical foundations of the fractal structure:} We next
describe how the self-similar fractal data can be described with
recently developed symbolic tools~\cite{Mitchell06}.  We first specify
a two-dimensional surface of section in the $xy p_x p_y$-phase space
by fixing the energy $E$ at -14.9 D$_2$ recoils and setting $y=0$.
Thus, every time a trajectory passes through the $x$-axis, we can
record $(x, p_x)$, defining a Poincar\'e map that maps a given
intersection $(x,p_x)$ forward to the next intersection $(x',p_x')$.
Fig.~\ref{fsos} shows the corresponding surface-of-section plot.  The
vertical line $\mathcal{L}_0$ at $x=0$ consists of all points at the
origin moving outward with arbitrary launch angle $\theta$.  The point
$\mathbf{z}_\mathsf{X}$ is the unstable fixed point equal to the
intersection of the unstable periodic orbit (PO) in
Fig.~\ref{fpotentials}b with the surface of section.  Attached to
$\mathbf{z}_\mathsf{X}$ are its stable $\mathcal{S}$ (thick) and
unstable $\mathcal{U}$ (thin) manifolds, consisting of all points that
asymptote to $\mathbf{z}_\mathsf{X}$ in the forward and backward
directions, respectively.  These manifolds intersect an infinite
number of times, forming an intricate pattern called a {\em homoclinic
tangle}~\cite{Wiggins92,Mitchell06,tangles,Mitchell03}.  The segments
of the stable and unstable manifolds connecting
$\mathbf{z}_\mathsf{X}$ to the point $\mathbf{P}_0$ define the shaded
region called the ``complex''.

Escape from the complex occurs via {\em escape lobes} $E_{n}$, defined
in Fig.~\ref{fsos} as the regions bounded by the stable and unstable
segments connecting $\mathbf{P}_{n}$ to $\mathbf{Q}_{n}$.  The lobe
$E_{-1}$, inside the complex, maps to $E_0$, outside the complex.
Once in $E_0$ a point then maps to $E_1$, $E_2$, etc., eventually
passing into the secondary well on the right.  The lobes $E_{-k}$
contain all points that escape in $k$ iterates.  These lobes become
progressively more stretched and folded as $k$ increases.  (An
analogous sequence of lobes $C_n$ controls \emph{capture} into the
complex.)  Note that we are able to chose physical parameters that
make the lobes quite large compared to Planck's constant
(Fig.~\ref{fsos}).

We plot as a function of $\theta$ the number of iterates for a point
to escape the complex, shown as the lower plot of each pair in
Fig.~\ref{ftemplates}.  These plots straighten each icicle into a
constant escape segment.  A segment that escapes on iterate $k$ is an
intersection between $E_{-k}$ and $\mathcal{L}_0$.  For example, the
segment at iterate two in Fig.~\ref{ftemplates}i is the intersection
with $E_{-2}$ in Fig.~\ref{fsos}.

Ref.~\cite{Mitchell06} introduces a symbolic technique, called {\em
homotopic lobe dynamics}, to compute the structure of escape segments
based on the tangle topology.  (See also Refs.~\cite{tangles}.)  We
summarize the results obtained from applying this technique to the
tangle in Fig.~\ref{fsos}.

The structure of the discrete-escape-time plot up to a given iterate
$n$ is specified by a string $\ell_n$ of symbols in the set $\{c_1,
c_2, a, b, f, u_0, u_1, u_2, ...\}$ as well as their inverses,
e.g. $c_1^{-1}$.  The first string is $\ell_1 = u_1 b^{-1} u_0$.  All
subsequent strings can be obtained from the first by mapping each
symbol forward according to the substitution rules:
\begin{subequations}
\begin{align}
c_1  &\mapsto c_2,  
&c_2 &\mapsto f^{-1} u_0 a u_0^{-1} f, \\
a   &\mapsto b^{-1} u_0^{-1} b, 
& f &\mapsto c_1^{-1} u_0^{-1} f, \\
u_n & \mapsto u_{n+1}, 
&b &\mapsto b^{-1} u_0^{-1} f,
\end{align}
\label{r6}
\end{subequations}
using the standard convention for iterating inverses, e.g. $b^{-1}
\mapsto f^{-1} u_0 b$.  For example, the first four strings are:
\begin{subequations}
\begin{align}
\ell_1 &= u_1 b^{-1} \underline{u_0}, 
\label{r2} \\
\ell_2 &= u_2 f^{-1} \underline{u_0} b u_1, 
\label{r3} \\
\ell_3 &= u_3 f^{-1} \underline{u_0} c_1 u_1 b^{-1} \underline{u_0^{-1}} f u_2,\label{r4} \\
\ell_4 &= u_4 f^{-1} \underline{u_0} c_1 u_1 c_2 u_2 f^{-1} \underline{u_0} b u_1^{-1} c_1^{-1} \underline{u_0^{-1}} f u_3.
\label{r5}
\end{align}
\label{r1}
\end{subequations}
An $\ell_n$ string encodes the discrete-escape-time plot as follows.
Each appearance of $u_0^{\pm 1}$ in $\ell_n$ (underlined for emphasis)
represents a segment that escapes at iterate $n$.  For example, the
$u_0$ factor in Eq.~(\ref{r2}) corresponds to the escape segment at
iterate one in Fig.~\ref{ftemplates}i.  This $u_0$ factor then maps
forward to $u_1$ in Eq.~(\ref{r3}).  In general, we see that each
$u_k^{\pm 1}$ in $\ell_n$ represents a segment that escapes at iterate
$n-k$.  In Eq.~(\ref{r3}), another $u_0$ factor has appeared,
corresponding to the escape segment at iterate two.  This segment is
to the left of the first segment, just as the $u_0$ factor in
Eq.~(\ref{r3}) is to the left of the $u_1$ factor.  In general, the
left-right ordering of $u_k^{\pm 1}$ symbols in $\ell_n$ represents
the left-right ordering of segments in the discrete-escape-time plot.
On the next two iterates, two new $u_0$ factors appear in
Eq.~(\ref{r4}), and three more appear in Eq.~(\ref{r5}), in agreement
with Fig.~\ref{ftemplates}i.

All other symbols besides $u_k^{\pm 1}$ represent gaps between
adjacent escape segments.  For example, $b^{-1}$ in Eq.~(\ref{r2})
represents the gap $b^{-1} (i)$ in Fig.~\ref{ftemplates}i, and
$b^{-1}$ in Eq.~(\ref{r4}) represents the gap $b^{-1} (ii)$ in
Figs.~\ref{ftemplates}i and ~\ref{ftemplates}ii.  The string $\ell_5$
will also contain a $b^{-1}$ factor, representing the gap $b^{-1}
(iii)$ in Figs.~\ref{ftemplates}ii and \ref{ftemplates}iii.  Since
each $b^{-1}$ factor generates exactly the same string of symbols
under Eqs.~(\ref{r6}), each gap labeled $b^{-1}$ in
Fig.~\ref{ftemplates} contains the same pattern of escape segments.
This means that $b^{-1}$ corresponds to a particular template, i.e. to
find occurrences of this template in the escape-time plot, we need
only look for occurrences of $b^{-1}$ in the expression for $\ell_n$.

It follows from the above logic that each symbol $\{c_1, c_2, a, b,
f\}$ generates its own template, with inverse symbols generating
reflected templates.  For example, mapping $f^{-1}$ forward three
times yields
\begin{equation}
f^{-1} u_0 c_1 u_1 c_2 u_2 f^{-1} u_0 a u_0^{-1} f.
\label{r10} 
\end{equation}
The reader may verify that Eq.~(\ref{r10}) describes the segments up
to iterates five, seven, and six in Figs.~\ref{ftemplates}a,
\ref{ftemplates}b, and \ref{ftemplates}c.
Note that different experimental parameters will yield different
algebraic rules and different templates.

This algebraic formalism computes a minimal set of escape segments,
but generally not all segments.  That is, at later times, we typically
find additional segments in the numerics.  This illustrates what has
previously been called an {\em epistrophic fractal}~\cite{Mitchell03}.
Nevertheless, unpredicted segments can be accommodated within an
updated algebraic formalism, as explained in Ref.~\cite{Mitchell06}.

{\em Conclusions:} We predict that experiments on the
intermediate-time escape dynamics of ultracold atoms from an optical
trap can directly image fractals.  The resolution is particularly good
when using a BEC.  The fractal structure depends on the topology of
homoclinic tangles, which are common to numerous chaotic systems.
Such experiments would thus provide a new laboratory tool for the
study of an important chaotic mechanism.  Similarly, an improved
understanding of the chaotic escape pathways of atoms from optical
traps could be relevant for the understanding of mixing and
thermalization in traps and for the control and coherent emission of
atomic wavepackets.  Finally, the dependence of these fractals on atom
density could serve as an interesting probe of atom-atom
interactions, a subject to be explored in future work.


\end{document}